\documentclass{elsart}

\usepackage{amsmath,amssymb,bm}
\usepackage{graphicx}
\usepackage{caption}
\usepackage{subcaption}
\usepackage[utf8]{inputenc}
\usepackage[english]{babel}

\begin{document}

\begin{frontmatter}

\title{A path to the Nuclear Equation of State within the frameworks of Mean-Field and Fermionic Dynamics}


\author{ T. Depastas$^{a}$, G.A. Souliotis$^{a,*}$, M. Veselsky$^{b}$, A. Bonasera$^{c,d}$ }


\address{ $^{a}$
           Laboratory of Physical Chemistry, Department of Chemistry,
           National and Kapodistrian University of Athens, Athens, Greece }
\address{ $^{b}$ Institute of Physics, Slovak Academy of
                     Sciences, Bratislava, Slovakia        }
\address{ $^{c}$ Cyclotron Institute, Texas A\&M University,
                     College Station, Texas, USA           }
\address{ $^{d}$ Laboratori Nazionali del Sud, INFN, Catania, Italy }

\address{ $^{*}$ Corresponding author. Email: soulioti@chem.uoa.gr }

\begin{abstract}
The nuclear Equation of State (EoS) lies in the center of the nuclear N-body problem, as it describes the properties of the Nuclear Matter (NM) and determines the parameters of the nuclear interaction.\\
In this work, we propose a theoretical description of the EoS of both Symmetric (SNM) and Asymmetric (ANM) nuclear matter within the framework of Fermionic Dynamics. With this description we produce several new semi-hard EoS's with density dependent effective mass.\\
Finally, we transform the aforementioned theory in order to be consistent with Mean-Field dynamics. We use this approach to accurately calculate the binding energies and charge radii of nuclei in the A=40--238 region with the Skyrme-Hartree-Fock (SHF) model.
\end{abstract}

\end{frontmatter}

\maketitle
\section{Introduction}
\label{s0}
\vspace{-0.8 cm}
The nuclear interaction that is responsible for the existence of bound nuclear systems, is one of the most complicated forces in nature. Although the description of the nuclear potential remains an open research topic, its origin can be considered to be the residual strong interaction between quarks and gluons inside the nucleons \cite{Takigawa}. There are various proposed theories that describe the nuclear interaction, ranging from Effective Field Theories (QCD) to realistic in-vacumn and in-medium potentials \cite{Ring}, as well as effective nuclear interactions with free parameters \cite{Sk1958}. Here, we use the latter approach and consequently, our studies aim at finding sets of parameters for the effective interactions that appropriately reproduce a large body of experimental data.\\
The Equation of State (EoS) of a nuclear system expresses the total energy per nucleon as a function of density and temperature \cite{Zheng2014} and is the key to constrain the parameters of the effective nuclear interaction. In addition, the EoS plays a central role in the description of the dynamical properties of both finite nuclei \cite{Zheng2014} and Nuclear Matter (NM) \cite{Papa2013} (i.e. a set of nucleons of macroscopic dimensions, without Coulomb and surface interactions). The EoS parameters are determined in part from the characteristics of Symmetric (SNM) and Asymmetric (ANM) Nuclear Matter, as can be encountered in astrophysical objects such as neutron stars.\\
The structure of this manuscript is as follows. First, in section \ref{s1} we present the Skyrme-like effective interactions for the Fermionic and Mean-Field Dynamics that we use later. In section \ref{s2}, we present a derivation of the SNM equations for the Constrained Molecular Dynamics (CoMD) functional, assuming a free nucleon mass. Then, in section \ref{s3}, we derive the SNM equations for the CoMD functional with a density dependent in-medium effective mass. The solutions of these SNM equations correspond to several EoS that are appropriate the CoMD model.\\
Furthermore, in section \ref{s4}, we extend our study to include the symmetry energy and develop an EoS for Asymmetric Nuclear Matter (ANM). We derive ANM equations with an appropriate functional form for the symmetry energy. With these results, we constrain the value of the asymmetry parameter ($\delta_x$) for NM into the range $\pm 0.72$. Additionally, we derive a complete EoS for NM, that can describe both symmetric and asymmetric systems.\\
Finally, in section \ref{s5}, we develop a procedure to transform the CoMD consistent EoS into a mean-field consistent EoS, using a Skyrme-Hartree-Fock (SHF) functional. With this process, we calculate the binding energies and charge radii for various nuclei in the range of $A=40-238$. Our EoS adequately reproduces the experimental data with a reduced Coulomb potential, due to Pauli correlations.
\section{The Effective Interaction and the EoS}
\label{s1}
\vspace{-0.8 cm}
The nucleon-nucleon interaction inside the nuclear medium can be described by the macroscopic properties of the Asymmetric and the Symmetric Nuclear Matter (ANM and SNM respectively). Here, we develop a unified EoS for the mean-field and Fermionic dynamics Skyrme Effective interactions \cite{Sk1958}. For the mean field interaction, we use the usual Skyrme-Hartree-Fock (SHF) interaction \cite{Vauth1971}, with 2-body and approximate 3-body terms. This is given by
\begin{equation}
    V_{SHF}=\sum_{i<j}v_{ij}+\sum_{i<j<k}v_{ijk}
    \label{e0}
\end{equation}
The two body $v_{12}$ and three body $v_{123}$ terms are given respectively by
\begin{multline}
v_{12}=t_{0}\left(1+x_{0} P_{\sigma}\right) \delta\left(\bm{\mathbf{r}}_{1}-\bm{\mathbf{r}}_{2}\right)
+\frac{1}{2} t_{1}\left[\delta\left(\bm{\mathbf{r}}_{1}-\bm{\mathbf{r}}_{2}\right) \bm{k}^{2}+\bm{k}^{\prime 2} \delta\left(\bm{\mathbf{r}}_{1}-\bm{\mathbf{r}}_{2}\right)\right] \\
+t_{2} \bm{\mathrm{k}}^{\prime} \cdot \delta\left(\bm{\mathbf{r}}_{1}-\bm{\mathbf{r}}_{2}\right) \bm{\mathrm{k}}+i W_{0}\left(\bm{\sigma}_{1}+\bm{\sigma}_{2}\right) \cdot \bm{\mathrm{k}}^{\prime} \times \delta\left(\bm{\mathbf{r}}_{1}-\bm{\mathbf{r}}_{2}\right) \bm{\mathrm{k}}
\label{e0a}
\end{multline}
\begin{multline}
    v_{123}=t_3 \delta\left(\bm{\mathbf{r}}_{1}-\bm{\mathbf{r}}_{2}\right) \delta\left(\bm{\mathbf{r}}_{1}-\bm{\mathbf{r}}_{3}\right) \delta\left(\bm{\mathbf{r}}_{2}-\bm{\mathbf{r}}_{3}\right) \\
    \approx \frac{t_3}{6} \left(1+P_{\sigma}\right) \delta\left(\bm{\mathbf{r}}_{1}-\bm{\mathbf{r}}_{2}\right) \rho^a\left(\frac{\bm{\mathbf{r}}_{1}+\bm{\mathbf{r}}_{2}}{2}\right)
    \label{e0b}
\end{multline}
where $t_i$, $i=1-3$ and $a$ are free parameters, $P_{\sigma}=\frac{1}{2}(1+\bm{\sigma}_{1}\cdot\bm{\sigma}_{2})$ is the spin-exchange operator and $\bm{k}=\frac{1}{2i}\left(\nabla_1-\nabla_2\right)=\bm{k}^{\prime\dag}$ is the wave-number operator for the relative motion. We stress that the dagger operator $\bm{k}^{\prime\dag}$ acts on the left, while the operator $\bm{k}$ acts on the right. The first term represents a volume contribution, the second and third simulate the symmetry, surface and effective mass contributions and the fourth term corresponds to the spin-orbit coupling. \\
For the fermionic dynamics interaction, we use the Skyrme-like potential of the Constrained Molecular Dynamics (CoMD) model \cite{Bon2001}. This effective interaction contains volume, 3-body, asymmetry, Coulomb and surface terms. Excluding the long range Coulomb interaction, the effective potential is expressed as a sum of contact terms
\begin{equation}
    V_{CoMD}=\sum_{i<j}v_{ij}=\sum_{i<j}v^{(2)}\delta(\bm{r_i}-\bm{r_j})
    \label{e0c}
\end{equation}
The two body local terms $v^{(2)}$ are given by
\begin{equation}
    v^{(2)}=\frac{T_0}{\rho_0}+\frac{2T_3\rho^{\sigma-1}}{(\sigma+1)\rho_0^{\sigma}}+\frac{a_{sym}}{\rho_0}\left(\frac{\rho}{\rho_0}\right)^{\gamma-1}\left(2\delta_{\tau_1,\tau_2}-1\right)
    +\frac{C_s}{\rho_0}\nabla^2_{\bm{R_1}}
    \label{e0d}
\end{equation}
where $T_0$, $T_3$, $a_{sym}$, $\sigma$, $C_s$ and $\gamma$ are free parameters, $\rho$ is the one-body density, $\rho_0$ is the saturation density of NM, $\tau_i$, $i=1,2$ are the isospins of each nucleon and $\bm{R_1}$ is the centroid of nucleon 1 in r-space, which corresponds to $\bm{R_1} = \langle \bm{r_1} \rangle$ \cite{Bon2001}. The first term of eq \ref{e0d} corresponds to the volume contribution, the second to an approximate 3-body contribution, the third to the symmetry potential and the last to the surface contribution.
\\
The total potential energy of NM in the context of Fermionic Dynamics is given then by the integral, as is discussed in \cite{Papa2013}
\begin{multline}
    W=\frac{1}{2}\int v_{12} f(\bm{r_1},\bm{p_1})f(\bm{r_2},\bm{p_2}) d^3\bm{r_1}d^3\bm{r_2}d^3\bm{p_1}d^3\bm{p_2} \\
    =\frac{1}{2}\int v^{(2)}\delta(\bm{r_1}-\bm{r_2}) \rho(\bm{r_1})\rho(\bm{r_2}) d^3\bm{r_1}d^3\bm{r_2}
    =\frac{1}{2}\int v^{(2)}\rho^2(\bm{r})d^3\bm{r}
    \label{e0e}
\end{multline}
where $f(\bm{r_i},\bm{p_i})$ is the 1-body phase-space distribution of each nucleon in the Wigner representation and $\rho(\bm{r_i}) = \int f(\bm{r_i},\bm{p_i}) d^3\bm{p_i}$ is the corresponding nucleonic density. The $\frac{1}{2}$ factor takes into account the two permutations of the indistinguishable pair of nucleons and avoids double counting. One can recognise the quantity of $\frac{1}{2}V^{(2)}\rho^2(\bm{r})$ as the potential energy density. The corresponding density within the mean-field framework, is defined as \cite{Vauth1971} $\rho(\bm{r})=\sum_{i,\sigma,\tau} |\phi_i(\bm{r},\sigma,\tau)|^2$, with $\phi_i(\bm{r},\sigma,\tau)$ the one-body wavefunction of a nucleon with spin $\sigma$ and isospin $\tau$. The total potential energy per nucleon of NM can be defined then as the energy per unit volume over the number of nucleons per unit volume, i.e.
\begin{equation}
    E^{pot}_{NM}=\frac{\frac{1}{2}v^{(2)}\rho^2(\bm{r})}{\rho(\bm{r})}=\frac{1}{2}v^{(2)}\rho
    \label{e0f}
\end{equation}
This potential energy forms the basis of the EoS in the NM limit. In the next sections, we derive the EoS of SNM and ANM for the Fermionic Dynamics effective interaction (eq \ref{e0a}) and we transform them into a form consistent with the mean-field interaction (eq \ref{e0d}).
\section{The SNM EoS of Fermionic Dynamics with free nucleon mass}
\label{s2}
\begin{figure}[!h]                                        
\centering
\includegraphics[width=12.0 cm]
{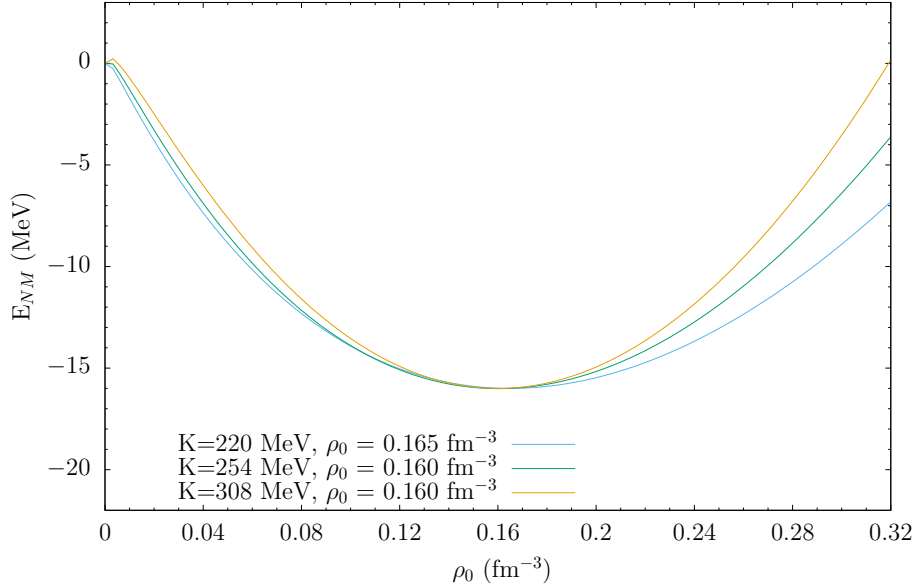} 
\caption{Equations of State without effective nucleonic mass. The compressibilities range from 220 MeV to 308 MeV, according to the key and the corresponding parameters are given in Table \ref{Tab1}.}
\label{Fig1}
\end{figure}

\vspace{-0.8 cm}
In the limit of SNM, where the Coulomb energy is neglected, only the volume and 3-body terms contribute to the total energy. Thus, the local 2-body potential is given only by the first two terms of eq \ref{e0d}. The potential part (according to eq \ref{e0f}) is then
\begin{equation}
    E^{pot}_{SNM}=\frac{T_0}{2}\frac{\rho}{\rho_0}+\frac{T_3}{\sigma+1}\left(\frac{\rho}{\rho_0}\right)^{\sigma}
    \label{e1}
\end{equation}
The kinetic part of the total energy of SNM per nucleon can be taken as the mean Fermi energy. First, will consider the mass of the nucleons to be constant and unaffected by the potential. This is a usual approximation in the literature (e.g. see \cite{Papa2013}). The kinetic part is then
\begin{equation}
    E^{kin}_{SNM}=\overline{\epsilon}_F=\frac{3}{5}\epsilon_F=\frac{3}{5}\frac{\hbar^2}{2m_0}\left(\frac{3\pi^2}{2}\rho\right)^{2/3}
    \label{e2}
\end{equation}
The potential and kinetic parts that we already described, give the total energy per nucleon of SNM
\begin{equation}
    E_{SNM}(\rho)=\frac{T_0}{2}\frac{\rho}{\rho_0}+\frac{T_3}{\sigma+1}\left(\frac{\rho}{\rho_0}\right)^{\sigma}+\overline{\epsilon}_F
    \label{e3}
\end{equation}
The known properties of nuclear matter require the calculation of the first two derivatives of the total energy with respect to the density, i.e. the EoS. First of all, the derivatives of the mean Fermi energy are
\begin{equation}
    \frac{\partial \overline{\epsilon}_F }{\partial \rho}=\frac{2}{3}\cdot\frac{3}{5}\frac{\hbar^2}{2m_0}\left(\frac{3\pi^2}{2}\right)^{2/3}\rho^{-1/3}=\frac{2\overline{\epsilon}_F}{3\rho}
    \label{e4}
\end{equation}
\begin{equation}
    \frac{\partial^2 \overline{\epsilon}_F }{\partial \rho^2}=-\frac{1}{3}\cdot\frac{2}{3}\cdot\frac{3}{5}\frac{\hbar^2}{2m_0}\left(\frac{3\pi^2}{2}\right)^{2/3}\rho^{-4/3}=-\frac{2\overline{\epsilon}_F}{9\rho^2}
    \label{e5}
\end{equation}
Then, the derivatives of the total energy per nucleon of SNM are 
\begin{equation}
    \frac{\partial E_{SNM} }{\partial \rho}=\frac{T_0}{2}\frac{1}{\rho_0}+\frac{\sigma}{\sigma+1}\frac{\rho^{\sigma-1}}{\rho_0^{\sigma}}T_3+\frac{2\overline{\epsilon}_F}{3\rho}
    \label{e6}
\end{equation}
\begin{equation}
    \frac{\partial^2 E_{SNM} }{\partial \rho^2}=\frac{\sigma(\sigma-1)}{\sigma+1}\frac{\rho^{\sigma-2}}{\rho_0^{\sigma}}T_3-\frac{2\overline{\epsilon}_F}{9\rho^2}
    \label{e7}
\end{equation}
In order to determine the parameters of the EoS, the usual properties of SNM at equilibrium $\rho=\rho_0$ will be used. The values of these properties are the usual used in the literature (e.g. \cite{Papa2013}). The equilibrium energy per nucleon of SNM is
\begin{equation}
    E_0\equiv E_{SNM}(\rho_0)=\frac{T_0}{2}+\frac{T_3}{\sigma+1}+\overline{\epsilon}^0_F=-BE_0/A
    \label{e8}
\end{equation}
Furthermore, the equilibrium state of SNM has the minimum energy. Equivalently the pressure of a Fermi gas with binding potential at $T=0$ in the equilibrium density is $0$, so
\begin{equation}
    P_0\equiv\rho_0\frac{\partial E_{SNM} (\rho_0) }{\partial \rho}=\frac{T_0}{2}+\frac{\sigma}{\sigma+1}T_3+\frac{2}{3}\overline{\epsilon}^0_F=0
    \label{e9}
\end{equation}
Finally, the compressibility of SNM is defined to be proportional to the curvature of the EoS at the equilibrium
\begin{equation}
    K_{NM}\equiv9\rho_0^2\frac{\partial^2 E_{SNM} (\rho_0) }{\partial \rho^2}=9T_3\frac{\sigma(\sigma-1)}{\sigma+1}-2\overline{\epsilon}^0_F
    \label{e10}
\end{equation}
where $\overline{\epsilon}^0_F=\frac{3}{5}\epsilon^0_F=\frac{3}{5}\frac{\hbar^2}{2m_0}\left(\frac{3\pi^2}{2}\rho_0\right)^{2/3}$ is the mean Fermi energy at the equilibrium density.\\
The SNM parameters of equilibrium density and energy per nucleon may be taken to be $\rho_0=0.165$ fm$^{-3}$ and $E_0=-16$ MeV. Then if the value of $\sigma$ is chosen, the constants $T_0$, $T_3$ and $K_{NM}$ can be found by the solution of the system of equations \ref{e8}-\ref{e10}. This approach is the one followed by Veselsky in private communication (code geneos.f). Here we will try to present a derivation of these equations. To that end, the equation \ref{e9} is rewritten as
\begin{equation}
    \frac{\sigma}{\sigma+1}T_3=-\frac{T_0}{2}-\frac{2}{3}\overline{\epsilon}^0_F
    \label{e11}
\end{equation}
Then by using equation \ref{e11} in \ref{e8}, we obtain:
\begin{equation}
    E_0=\frac{T_0}{2}+\frac{1}{\sigma}\left( -\frac{T_0}{2}-\frac{2}{3}\overline{\epsilon}^0_F\right)+\overline{\epsilon}^0_F=\frac{T_0}{2}\left(1-\frac{1}{\sigma}\right)+\overline{\epsilon}^0_F\left(1-\frac{2}{3\sigma}\right)
    \label{e12}
\end{equation}
from which we obtain $T_0$ as:
\begin{equation}
    T_0=2\frac{\sigma E_0-\overline{\epsilon}^0_F\left(\sigma-\frac{2}{3}\right)}{\sigma-1}
    \label{e13}
\end{equation}
The value of $K_{NM}$ is determined by equation \ref{e10}, if $T_3$ is determined. This constant can be determined equivalently from equations 19 \ref{e11} and \ref{e13}
\begin{equation}
    T_3=-\frac{\sigma+1}{\sigma}\left(\frac{T_0}{2}+\frac{2}{3}\overline{\epsilon}^0_F\right)=\frac{\sigma+1}{\sigma-1}\left(\frac{1}{3}\overline{\epsilon}^0_F-E_0\right)
    \label{e14}
\end{equation}
An alternative approach to the problem of determining the EoS would be the calculation of $T_0$, $T_3$ and $\sigma$, given the values of $\rho_0$, $E_0$ and $K_{NM}$. This procedure is followed in \cite{Papa2013}. The values of $T_0$, $T_3$ are calculated by equations 21-22 if $\sigma$ is determined. This can be done with the use of equations \ref{e13} and \ref{e14}
\begin{equation}
      K_{NM}=9T_3\frac{\sigma(\sigma-1)}{\sigma+1}-2\overline{\epsilon}^0_F= 9\frac{\sigma+1}{\sigma-1}\left(\frac{1}{3}\overline{\epsilon}^0_F-E_0\right)\frac{\sigma(\sigma-1)}{\sigma+1}-2\overline{\epsilon}^0_F
      \label{e15}
\end{equation}
or equivalently 
\begin{equation}
    \sigma=\frac{K_{NM}+2\overline{\epsilon}^0_F}{9\left(\frac{1}{3}\overline{\epsilon}^0_F-E_0\right)}
    \label{e16}
\end{equation}
In order to calculate the potential parameters and the corresponding EoS, we wrote a simple code based upon the aforementioned equations. The results of $T_3$, $T_0$ and $\sigma$ with certain values of $K_{NM}$, $\rho_0$ and $E_0$ can be seen in Table \ref{Tab1}. The $\rho_0=0.160$ fm$^{-3}$ EoS were constructed from Veselsky and were used in our previous work \cite{Teo2021}.
\begin{table}[]
\caption{EoS parameters without effective mass, $m^*=m$.}
\label{Tab1}
\begin{tabular}{|l|l|l|l|l|l|}
\hline
\textbf{K$_{NM}$ (MeV)} & \textbf{$\rho_0$ (fm$^{-3}$)} & \textbf{-$E_0$ (MeV)} & \textbf{$T_0$ (MeV)} & \textbf{$T_3$ (MeV)} & \textbf{$\sigma$} \\ \hline
220                     & 0.165                         & 16                    & -266.919             & 212.734              & 1.250             \\ \hline
254                     & 0.160                         & 16                    & -190.250             & 136.800              & 1.414             \\ \hline
308                     & 0.160                         & 16                    & -147.010             & 93.600               & 1.670             \\ \hline
\end{tabular}
\end{table}
The three aforementioned EoS can be plotted as the total energy per nucleon of SNM with respect to the density, presented in Figure \ref{Fig1}, for compressibility values according to the key. We observe that the 'hard' EoS with $K_{NM}=308$ MeV has a local maximum at very low densities indicating that the system is unbound. This is an interesting feature that may be connected with clusterization, as it is discussed in the next section.

\section{The SNM EoS of Fermionic Dynamics with Effective Mass}
\label{s3}
\begin{figure}[!h]                                        
\centering
\includegraphics[width=12.0 cm]
{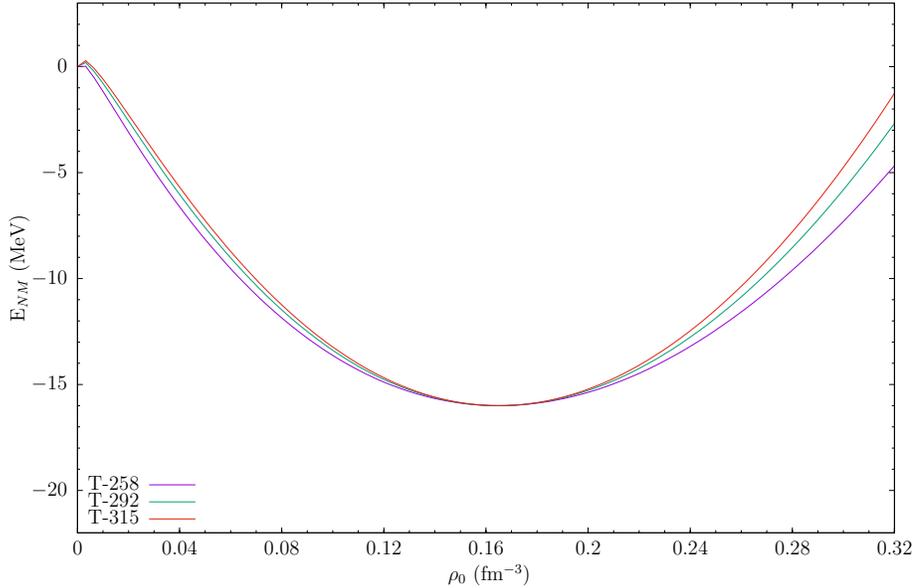} 
\caption{Equations of State with effective mass. The compressibilities range from 220 MeV to 275 MeV, according to the key and the corresponding parameters are given in Table \ref{Tab2}.}
\label{Fig2}
\end{figure}

In order to simulate the finite range of the nuclear interaction the potential must contain momentum dependent terms. This is the basic idea behind the low energy expansion of the Skyrme's potential \cite{Sk1958}. The momentum dependent term that could be added to the CoMD potential, should vanish at high nucleon momenta or low densities. This characteristic of the potential ensures the consistency with energy conservation in a nucleon ejection event. Thus the 1-body momentum dependent term could have the form
\begin{equation}
    V_i^m = -T_m e^{-\alpha \bm{P_i}^2}
    \label{e17}
\end{equation}
In the low energy limit this term takes the form 
\begin{equation}
V_i^m \rightarrow -T_m (1-\alpha \bm{P_i}^2)
\label{e18}
\end{equation}
If we neglect the constant $T_m$ term from the previous equation, the momentum dependent potential contributes to the kinetic energy of the CoMD model as 
\begin{equation}
T=\frac{\bm{P_i}^2}{2m_0} \rightarrow \frac{\bm{P_i}^2}{2m_0}+T_m\alpha \bm{P_i}^2 \equiv \frac{\bm{P_i}^2}{2m^*}
\label{e19}
\end{equation}
where $m^*$ is the effective mass of the nucleon inside the nuclear medium. If we define the effective mass ratio parameter as $\frac{1}{f} \equiv \frac{m_0}{m^*}$, then this parameter has the form 
\begin{equation}
\frac{1}{f}=1+2m_0T_m\alpha
\label{e20}
\end{equation}
The effective mass parameter in the usual Skyrme formulation has a density dependence and can be written as \cite{Cochet2004}
\begin{equation}
    \frac{1}{f(\rho)}=1+\frac{1}{8}\frac{m_0}{\hbar^2}\rho\Theta_s \equiv 1 +\lambda \rho
    \label{e21}
\end{equation}
where $\lambda \equiv \frac{1}{8}\frac{m_0}{\hbar^2}\Theta_s$ and $\Theta_s$ contains various of the usual coefficients of a Skyrme potential, i.e. $\Theta_s=3t_1+(5+4x_2)t_2$. In our previous work \cite{Teo2021} we used the effective mass parameter $f$ as a free variable, completely decoupled from the $K_{NM}$ in order to study its effect. Here, by choosing a certain form for the momentum dependence of the effective mass, we develop a model that describes the EoS with an effective mass consistent with the compressibility and the other SNM characteristics. To that end, we rewrite the equations presented in the previous section with the transformation
\begin{equation}
\overline{\epsilon}_F \rightarrow \overline{\epsilon}_F\frac{1}{f}=\overline{\epsilon}_F(1 +\lambda \rho)
\label{e22}
\end{equation}
This transformation implies that the total energy per nucleon of SNM takes the form
\begin{equation}
     E_{SNM}(\rho)=\frac{T_0}{2}\frac{\rho}{\rho_0}+\frac{T_3}{\sigma+1}\left(\frac{\rho}{\rho_0}\right)^{\sigma}+\overline{\epsilon}_F\frac{1}{f}
     \label{e23}
\end{equation}
Additionally, the first derivative of the scaled mean Fermi energy is rewritten as
\begin{equation}
    \frac{\partial ( \overline{\epsilon}_F\frac{1}{f} ) }{\partial \rho} (\rho)=\frac{\partial \overline{\epsilon}_F}{\partial \rho}(\rho)\frac{1}{f} + \overline{\epsilon}_F\frac{\partial}{\partial \rho}(1 +\lambda \rho)=\frac{2\overline{\epsilon}_F}{3\rho}\frac{1}{f}+\overline{\epsilon}_F\lambda =\frac{\overline{\epsilon}_F}{\rho}\left( \frac{2}{3f} + \lambda\rho\right)
    \label{e23a}
\end{equation}
Equation \ref{e21} was used to achieve the above result. The last term in the above relation can be written as a function of $f$ as $\lambda\rho=\frac{1}{f}-1$. Thus, the above equation takes the form
\begin{equation}
    \frac{\partial ( \overline{\epsilon}_F\frac{1}{f} ) }{\partial \rho} (\rho)=\frac{\overline{\epsilon}_F}{\rho}\left( \frac{2}{3f} + \frac{1}{f}-1\right)=\frac{\overline{\epsilon}_F}{\rho}\left( \frac{5}{3f}-1\right)
    \label{e24}
\end{equation}
The kinetic energy term in the SNM equation for the pressure (i.e. eq \ref{e2}) becomes
\begin{equation}
    \rho_0\frac{\partial ( \overline{\epsilon}_F\frac{1}{f} ) }{\partial \rho}(\rho_0)=\overline{\epsilon}^0_F \left( \frac{5}{3f}-1 \right)
    \label{e25}
\end{equation}
In the same manner we calculate the additional term in the second derivative of the kinetic energy term. The second derivative of the scaled Fermi energy is
\begin{equation}
  \frac{\partial^2 ( \overline{\epsilon}_F\frac{1}{f} ) }{\partial \rho^2} (\rho) = \frac{\partial}{\partial \rho}(\frac{\partial}{\partial \rho}(\overline{\epsilon}_F\frac{1}{f})=\frac{\partial}{\partial \rho}(\frac{\partial\overline{\epsilon}_F }{\partial \rho}\frac{1}{f}+\overline{\epsilon}_F\frac{\partial \frac{1}{f} }{\partial \rho})
  \label{e26}
\end{equation}
or equivalently 
\begin{equation}
  \frac{\partial^2 ( \overline{\epsilon}_F\frac{1}{f} ) }{\partial \rho^2} (\rho) = \frac{\partial^2 \overline{\epsilon}_F}{\partial \rho^2}(\rho)\frac{1}{f} + \frac{\partial\overline{\epsilon}_F }{\partial \rho}(\rho)\frac{\partial}{\partial \rho}(1 +\lambda \rho)+\frac{\partial\overline{\epsilon}_F }{\partial \rho}(\rho)\lambda= \frac{\partial^2 \overline{\epsilon}_F}{\partial \rho^2}(\rho)\frac{1}{f}+2\lambda\frac{\partial\overline{\epsilon}_F }{\partial \rho}(\rho)
  \label{e27}
\end{equation}
Using equations \ref{e21} and \ref{e22}, this term becomes
\begin{equation}
    \frac{\partial^2 ( \overline{\epsilon}_F\frac{1}{f} ) }{\partial \rho^2} (\rho) =-\frac{2\overline{\epsilon}_F}{9\rho^2}\frac{1}{f}+2\lambda\frac{2\overline{\epsilon}_F}{3\rho}=\frac{\overline{\epsilon}_F}{9\rho^2}(-\frac{2}{f}+12\lambda\rho)
    \label{e28}
\end{equation}
or equivalently
\begin{equation}
    \frac{\partial^2 ( \overline{\epsilon}_F\frac{1}{f} ) }{\partial \rho^2} (\rho) =\frac{\overline{\epsilon}_F}{9\rho^2}(-\frac{2}{f}+\frac{12}{f}-12)=\frac{\overline{\epsilon}_F}{9\rho^2}(\frac{10}{f}-12)
    \label{e29}
\end{equation}
The kinetic energy term in the SNM equation for the compressibility (i.e. eq \ref{e10}) then becomes
\begin{equation}
     9\rho_0^2\frac{\partial^2 ( \overline{\epsilon}_F\frac{1}{f} ) }{\partial \rho^2} (\rho_0) =\left( \frac{10}{f} -12 \right)\overline{\epsilon}^0_F
     \label{e30}
\end{equation}
Returning to equations \ref{e8}, \ref{e9} and \ref{e10} and including the new terms from equations \ref{e22}, \ref{e25} and \ref{e30} respectively that have taken into account the effective mass and its derivatives, the SNM equations take the form
\begin{equation}
    E_0=\frac{T_0}{2}+\frac{T_3}{\sigma+1}+\overline{\epsilon}^0_F\frac{1}{f}=-BE_0/A
     \label{e31}
\end{equation}
\begin{equation}
    P_0=\frac{T_0}{2}+\frac{\sigma}{\sigma+1}T_3+\left( \frac{5}{3f}-1 \right)\overline{\epsilon}^0_F=0
    \label{e32}
\end{equation}
\begin{equation}
    K_{NM}=9T_3\frac{\sigma(\sigma-1)}{\sigma+1}+\overline{\epsilon}^0_F\left( \frac{10}{f} -12 \right)
    \label{e33}
\end{equation}
We stress that in equations \ref{e31}-\ref{e33} the effective mass function is evaluated at $\rho=\rho_0$. The validity of these equations can be checked by testing them with $f=1$. Indeed with effective mass value of 1 they reduce to equations \ref{e8}-\ref{e10}. For the determination of the potential parameters, there are three equations (\ref{e31}-\ref{e33}) and four unknown variables $T_0$, $T_3$, $\sigma$ and $K_{NM}$ or $f$. In order to solve this problem, we freely select the value of $\sigma$ and solve for the $T_0$, $T_3$, and $f$ with $\rho_0$, $E_0$ and $K_{NM}$ as input variables. This process for the mean-field Skyrme interaction is illustrated in \cite{Cochet2004}, \cite{Vauth1971}. The $\sigma$ values are chosen to give reasonable results for the SNM characteristics. Then, by subtracting eq \ref{e32} from \ref{e31} we obtain:
\begin{equation}
    E_0=T_3 \frac{1-\sigma}{1+\sigma}+\overline{\epsilon}^0_F\left( \frac{1}{f}-\frac{5}{3f}+1\right)\Rightarrow E_0=T_3 \frac{1-\sigma}{1+\sigma}+\overline{\epsilon}^0_F\left( 1-\frac{2}{3f}\right)
    \label{e34}
\end{equation}
from which we obtain for $T_3$:
\begin{equation}
    T_3=\frac{1+\sigma}{1-\sigma}\left[\overline{\epsilon}^0_F\left( 1-\frac{2}{3f}\right) - E_0 \right]
    \label{e35}
\end{equation}
Then equation \ref{e35} can be inserted to equation \ref{e33} to give:
\begin{multline}
    K_{NM}=9\sigma\left[\overline{\epsilon}^0_F\left( 1-\frac{2}{3f}\right) - E_0 \right]+\overline{\epsilon}^0_F\left( \frac{10}{f} -12 \right)\\
    =\frac{2\overline{\epsilon}^0_F}{f}(5-3\sigma)+3\overline{\epsilon}^0_F(3\sigma-4)-9\sigma E_0
    \label{e36}
\end{multline}
from which $f$ can be obtained as:
\begin{equation}
    \frac{1}{f}=\frac{1}{2(5-3\sigma)\overline{\epsilon}^0_F}\left[K_{NM}-3\overline{\epsilon}^0_F(3\sigma-4)+9\sigma E_0 \right]
    \label{e37}
\end{equation}
From eq \ref{e35} we get
\begin{equation}
 \frac{T_3}{1+\sigma}=\frac{1}{1-\sigma}\left[\overline{\epsilon}^0_F\left( 1-\frac{2}{3f}\right) - E_0 \right]  
 \label{e38}
\end{equation}
this is substituted in eq \ref{e32} to give
\begin{equation}
    \frac{T_0}{2}+\sigma\left( E_0 -\frac{T_0}{2}-\frac{\overline{\epsilon}^0_F}{f} \right)+\left( \frac{5}{3f}-1 \right)\overline{\epsilon}^0_F=0
    \label{e39}
\end{equation}
or equivalently 
\begin{equation}
    \frac{T_0}{2}(\sigma-1)=\sigma E_0 + \frac{\overline{\epsilon}^0_F}{f} \left( \frac{5}{3}-\sigma\right) - \overline{\epsilon}^0_F
    \label{e40}
\end{equation}
from which $T_0$ is obtained as:
\begin{equation}
    T_0=\frac{2}{\sigma-1}\left[\sigma E_0 + \frac{\overline{\epsilon}^0_F}{f} \left( \frac{5}{3}-\sigma\right) - \overline{\epsilon}^0_F\right]
    \label{e41}
\end{equation}
Equations \ref{e35}, \ref{e36} and \ref{e41} reduce to the corresponding equations without effective mass presented in the previous section, if we set $f=1$. Alternatively, we can calculate the parameters $T_0$, $T_3$ and $K_{NM}$, given the values of $f$, $\rho_0$, $\sigma$ and $E_0$ as input, by equations \ref{e41}, \ref{e35} and \ref{e36}, respectively. \\
To solve the SNM equations with effective mass, we have written a code that gives the potential parameters when provided with different input sets. One of its functions is to calculate the values of $T_0$, $T_3$ and $K_{NM}$  with $f$, $E_0$, $\rho_0$ and $\sigma$ as input. Another function is to calculate the values of $T_0$, $T_3$ and $f$, with $\sigma$, $E_0$, $\rho_0$ and $K_{NM}$ as input. Here we present some new EoS as the solutions of the SNM equations. The corresponding parameters can be found in Table \ref{Tab2}. With the value of $f=\frac{m^*}{m}=0.85$ at $\rho=\rho_0$ and the assumed linear density dependence of eq \ref{e21}, we can determine the slope $\lambda=1.07$. Thus we can write
\begin{equation}
    \frac{1}{f}=1+\lambda\rho=1+\left(\lambda\rho_0\right)\left(\frac{\rho}{\rho_0}\right) \Rightarrow  \frac{1}{f}=1+1.07\left(\frac{\rho}{\rho_0}\right)
    \label{e42}
\end{equation}
We insert this in eq \ref{e31} and along with the other determined parameters we plot the indicated EoS's in Figure \ref{Fig2}.
\begin{table}[]
\caption{EoS parameters with effective mass, $m^*<m$.}
\label{Tab2}
\begin{tabular}{|l|l|l|l|l|l|l|l|}
\hline
\textbf{EoS} & \textbf{K$_{NM}$ (MeV)} & \textbf{$\rho_0$ (fm$^{-3}$)} & \textbf{-$E_0$ (MeV)} & \textbf{$T_0$ (MeV)} & \textbf{$T_3$ (MeV)} & \textbf{$\sigma$} & \textbf{m$^*$/m} \\ \hline
T-258      & 258                     & 0.165                         & 16                    & -190.617             & 125.650               & 1.40              & 0.85             \\ \hline
T-292      & 292                     & 0.165                         & 16                    & -158.121             & 93.154               & 1.58              & 0.85             \\ \hline
T-315      & 315                     & 0.165                         & 16                    & -145.742             & 80.365               & 1.70              & 0.85             \\ \hline
\end{tabular}
\end{table}

From the previous equations and the parameters of the new EoS we confirm the strong correlation of $f$ and $K_{NM}$, as also discussed in \cite{Zheng2014}. In figure \ref{Fig2}, again we note that the system is unbound at very low densities. This may indicate the occurrence of clusterization at very low density. This peak can be used to justify a 'phase-transition' in the low density SNM in the spirit of Maxwell construction of a Van der Waals gas. This phase transition of course, would lead to a nuclear state were the nucleons would transform into a 'Fermi liquid'. This state corresponds to the formation of nucleonic clusters. The new EoS are used to simulate the Giant Dipole and Monopole Resonances (GDR and GMR respectively) with the CoMD model. These results are reported in \cite{??}.
\label{s4}
\section{The ANM EoS of Fermionic Dynamics with Effective Mass}
The EoS of Asymmetric Nuclear Matter (ANM) has, in addition to the terms described in the previous section, contributions from the kinetic and potential symmetry energy. These terms stem from the fact that a neutron-proton pair interacts more strongly than a neutron-neutron or a proton pair. The differences in the interaction energy may be intuitively understood in the spirit of the Yukawa theory, where the $np$ pair is able to exchange the $\pi^{\pm}$ pion which is heavier than the $\pi^0$ pion exchanged in $nn$ and $pp$ interactions. Then, according to the Uncertainty Principle, the $np$ interaction should have a shorter range than the $nn$, $pp$ interactions and, thus, it is stronger. Furthermore, part of the symmetry energy is considered as Pauli blocking, because the different N and Z numbers have a different fermi energy. The difference of their Fermi energies corresponds to the symmetry energy.\\
In the contex of the Fermionic Dynamics CoMD model, the potential symmetry energy is taken into account by the third term of eq \ref{e0d} that takes the form 
\begin{equation}
    \frac{a_{sym}}{2}\left(\frac{\rho}{\rho_0}\right)^{\gamma}\delta^2_x
    \label{es0a}
\end{equation}
The kinetic part of the symmetry energy comes from a two component asymmetric Fermi Gas \cite{}. This term can be expressed as  
\begin{equation}
    \frac{5}{9}\frac{\overline{\epsilon}_F}{f}\delta^2_x
    \label{es0b}
\end{equation}
Then, the total EoS of ANM is then given by  \cite{Papa2013}
\begin{equation}
     E_{ANM}(\rho)=\frac{T_0}{2}\frac{\rho}{\rho_0}+\frac{T_3}{\sigma+1}\left(\frac{\rho}{\rho_0}\right)^{\sigma}+\frac{a_{sym}}{2}\left(\frac{\rho}{\rho_0}\right)^{\gamma}\delta^2_x+\overline{\epsilon}_F\frac{1}{f}\left(1+\frac{5}{9}\delta^2_x\right)
     \label{es1}
\end{equation}
The variable $\delta_x=\frac{\rho_n-\rho_p}{\rho}=\frac{N-Z}{A}$ is the isospin asymmetry parameter and $\rho_n$, $\rho_p$ are the neutron and proton densities respectively.\\
The asymmetry parameter ($\delta_x$) is related to the more familiar total nuclear isospin's z-projection $\tau_3=\frac{N-Z}{2}$ and the neutron to proton ratio $N/Z$. These relations are derived according to the following procedure.
\begin{equation}
    \delta_x=\frac{N-Z}{A}=\frac{N-Z}{2}\frac{2}{A}=\tau_3\frac{2}{A}
    \label{es2}
\end{equation}
or equivalently
\begin{equation}
    \tau_3= \frac{A}{2} \delta_x
    \label{es3}
\end{equation}
Additionally, 
\begin{equation}
    \delta_x=\frac{N-Z}{A}=\frac{2N-A}{A}=2\frac{N}{A}-1=\frac{2}{1+\frac{Z}{N}}-1
    \label{es4}
\end{equation}
Then solving for $\frac{Z}{N}$
\begin{equation}
    \frac{Z}{N}=\frac{2}{\delta_x+1}-1=\frac{2-\delta_x-1}{\delta_x+1}=\frac{1-\delta_x}{\delta_x+1}
    \label{es5}
\end{equation}
or equivalently,
\begin{equation}
    \frac{N}{Z}=\frac{1+\delta_x}{1-\delta_x}
    \label{es6}
\end{equation}
With the use of eq \ref{es1} and the isospin asymmetry parameter $\delta_x$, we can express $E_{ANM}(\rho)$ approximately (up to the second power of the asymmetry parameter) as:
\begin{equation}
    E_{ANM}(\rho)=E_{SNM}(\rho)+S(\rho)\delta^2_x
    \label{es6a}
\end{equation}
with $E_{SNM}(\rho)$ as in eq \ref{e1}, \ref{e2} and \ref{e22}, while $S(\rho)$ is given by \cite{Papa2013}:
\begin{equation}
    S(\rho)\equiv\frac{\partial^2}{\partial \delta^2_x } E_{ANM}=\frac{5}{9}\overline{\epsilon}_F\frac{1}{f} + \frac{a_{sym}}{2}\left(\frac{\rho}{\rho_0}\right)^{\gamma}
    \label{es7}
\end{equation}
Furthermore, we define the quantities $L(\rho)$ and $K_{sym}(\rho)$ which are the slope of the symmetry energy and the symmetry energy contribution to the compressibility, respectively. These are defined similarly to the Pressure and Compressibility (eq \ref{e9}, \ref{e10}):
\begin{equation}
    L(\rho)\equiv3\rho\frac{\partial S}{\partial \rho}=3\rho\frac{5}{9}\frac{\partial ( \overline{\epsilon}_F\frac{1}{f} ) }{\partial \rho}+3\rho\frac{a_{sym}}{2}\frac{\partial}{\partial \rho}\left(\frac{\rho}{\rho_0}\right)^{\gamma}
    \label{es8}
\end{equation}
\begin{equation}
    K_{sym}(\rho)\equiv 9\rho^2 \frac{\partial^2 S}{\partial \rho^2}=9\rho^2\frac{5}{9}\frac{\partial^2 ( \overline{\epsilon}_F\frac{1}{f} ) }{\partial \rho^2}+9\rho^2\frac{a_{sym}}{2}\frac{\partial^2}{\partial \rho^2}\left(\frac{\rho}{\rho_0}\right)^{\gamma}
    \label{es9}
\end{equation}
Using the derivatives of the Fermi energy from eq \ref{e24}, \ref{e29}, the aforementioned quantities are written as follows
\begin{equation}
    L(\rho)=\frac{15\rho}{9}\frac{\overline{\epsilon}_F}{\rho}\left( \frac{5}{3f}-1\right)+3\rho\gamma\frac{a_{sym}}{2}\frac{\rho^{\gamma-1}}{\rho_0^{\gamma}}
    \label{es10}
\end{equation}
\begin{equation}
  K_{sym}(\rho)=\frac{45\rho^2}{9}\frac{\overline{\epsilon}_F}{9\rho^2}(\frac{10}{f}-12)+\gamma(\gamma-1)\frac{9\rho^2a_{sym}}{2}\frac{\rho^{\gamma-2}}{\rho_0^{\gamma}}
  \label{es11}
\end{equation}
or equivalently
\begin{equation}
    L(\rho)=\frac{5\overline{\epsilon}_F}{3}\left( \frac{5}{3f}-1\right)+\gamma\frac{3 a_{sym}}{2}\left(\frac{\rho}{\rho_0}\right)^{\gamma}
    \label{es12}
\end{equation}
\begin{equation}
  K_{sym}(\rho)=\frac{5\overline{\epsilon}_F}{9}(\frac{10}{f}-12)+\gamma(\gamma-1)\frac{9a_{sym}}{2}\left(\frac{\rho}{\rho_0}\right)^{\gamma}
  \label{es13}
\end{equation}
In the equilibrium state of ANM, the density is equal to the saturation density $\rho_0$ and eq \ref{es7}, \ref{es12} and \ref{es13} are written as:
\begin{equation}
    S_0\equiv S(\rho_0)=\frac{5}{9}\overline{\epsilon}_F\frac{1}{f} + \frac{a_{sym}}{2}
    \label{es14}
\end{equation}
\begin{equation}
    L_0\equiv L(\rho_0)=\frac{5\overline{\epsilon}^0_F}{3}\left( \frac{5}{3f}-1\right)+\frac{3 a_{sym}}{2}\gamma
    \label{es15}
\end{equation}
\begin{equation}
  K_{sym}^0\equiv K_{sym}(\rho_0)=\frac{5\overline{\epsilon}^0_F}{9}(\frac{10}{f}-12)+\frac{9a_{sym}}{2}\gamma(\gamma-1)
  \label{es16}
\end{equation}
Usually, we determine the parameters $a_{sym}$ and $\gamma$ such as they reproduce the expected empirical values of $S_0$, $L_0$ and $K^0_{sym}$ \cite{Zheng2014}. From eq \ref{es14} we obtain $a_{sym}$
\begin{equation}
    \frac{a_{sym}}{2}=S_0-\frac{5}{9}\overline{\epsilon}_F\frac{1}{f}
    \label{es16a}
\end{equation}
or equivalently 
\begin{equation}
    a_{sym}=2\left(S_0-\frac{5}{9}\overline{\epsilon}_F\frac{1}{f}\right)
    \label{es16b}
\end{equation}
By substituting $a_{sym}$ from eq \ref{es16a} into eq \ref{es15}, we can determine the value of $\gamma$ as follows:
\begin{equation}
    L_0=\frac{5\overline{\epsilon}^0_F}{3}\left( \frac{5}{3f}-1\right)+3\gamma\left(S_0-\frac{5}{9}\overline{\epsilon}_F\frac{1}{f}\right)
    \label{es16c}
\end{equation}
or equivalently
\begin{equation}
    3\gamma\left(S_0-\frac{5}{9}\overline{\epsilon}_F\frac{1}{f}\right)=L_0-\frac{5\overline{\epsilon}^0_F}{3}\left( \frac{5}{3f}-1\right)
    \label{es16da}
\end{equation}
and finally, by solving for $\gamma$ we get
\begin{equation}
    \gamma=\frac{L_0-\frac{5\overline{\epsilon}^0_F}{3}\left( \frac{5}{3f}-1\right)}{3\left(S_0-\frac{5}{9}\overline{\epsilon}_F\frac{1}{f}\right)}
    \label{es16d}
\end{equation}
Using eq \ref{es16b} and \ref{es16d} we can determine the values of the $a_{sym}$ and $\gamma$ parameters, if $S_0$, $L_0$ and $K^0_{sym}$ are known. According to \cite{Papa2013,Zheng2014}, $S_0=28.6$ MeV and $L_0=50\pm 40$ MeV. Alternatively, with the use of eq \ref{es14}, \ref{es15} and \ref{es16} and known values of $a_{sym}$ and $\gamma$, we can determine the aforementioned quantities of ANM. Here, we use an EoS with $\rho_0=0.165$ fm$^{-3}$, $a_{sym}=32$ MeV and $\gamma=1$ and, thus, the properties of ANM are $S_0=30.87$ MeV, $L_0=84.43$ MeV and $K_{sym}^0=-2.97$ MeV. Consequently, the total EoS of ANM (from eq \ref{es1}), for $E_0=-16.0$ MeV is 
\begin{equation}
    E_{ANM}^0\equiv E_{ANM}\left(\rho_0\right)=-16.0 +30.87\delta^2_x \; MeV
    \label{es17}
\end{equation}
From the requirement of the system to be bound $\left( E_{ANM}^0 \leq 0 \right)$, we may estimate the extreme values of the isospin asymmetry parameter and the relevant quantities $\tau_3$ and $N/Z$ (or $N/A$ equivalently).
\begin{equation}
    E_{ANM}^0 \leq 0 \rightarrow \delta^2_x \leq \frac{16.0}{30.87}=0.52
    \label{es18}
\end{equation}
or equivalently 
\begin{equation}
    -0.72\leq \delta_x \leq 0.72
    \label{es19}
\end{equation}
By substituting the above result into eq \ref{es3} and \ref{es6}, we get
\begin{equation}
    -0.36 \leq \frac{\tau_3}{A} \leq 0.36
    \label{es20}
\end{equation}
\begin{equation}
    0.16 \leq \frac{N}{Z} \leq 6.14
    \label{es21}
\end{equation}
The aforementioned results, of course, do not correspond to finite systems with Coulomb and Surface interactions, but they represent extreme boundary values for extended systems such as neutron stars or hyper-heavy nuclei.\\
The EoS for ANM of eq \ref{es17} can be generalized for any density, with the use of eq \ref{es1} and the definitions of the total pressure and compressility of ANM:
\begin{equation}
    P_{ANM} (\rho) \equiv \rho\frac{\partial E_{ANM}}{\partial \rho }=\rho\frac{\partial E_{SNM}}{\partial \rho }+\rho\frac{\partial S}{\partial \rho }\delta^2_x=P_{SNM}+\delta^2_x\frac{L}{3}
    \label{es22}
\end{equation}
\begin{equation}
    K_{ANM} (\rho) \equiv 9\rho^2\frac{\partial^2 E_{ANM}}{\partial \rho^2 }=9\rho^2\frac{\partial^2 E_{SNM}}{\partial \rho^2 }+9\rho^2\frac{\partial^2 S}{\partial \rho^2 }\delta^2_x=K_{SNM}+\delta^2_x K_{sym}
    \label{es23}
\end{equation}
By substituting eq \ref{e32}, \ref{e33} into the above results, eq \ref{es22} and \ref{es23} are rewritten as
\begin{equation}
    P_{ANM} (\rho)=\frac{T_0}{2}\left(\frac{\rho}{\rho_0}\right)+\frac{\sigma}{\sigma+1}T_3\left(\frac{\rho}{\rho_0}\right)^{\sigma}+\left( \frac{5}{3f}-1 \right)\overline{\epsilon}_F+\delta^2_x\frac{L}{3}
    \label{es24}
\end{equation}
\begin{equation}
    K_{ANM} (\rho)=9T_3\frac{\sigma(\sigma-1)}{\sigma+1}\left(\frac{\rho}{\rho_0}\right)^{\sigma}+\overline{\epsilon}_F\left( \frac{10}{f} -12 \right)+\delta^2_x K_{sym}
    \label{es25}
\end{equation}
Subtracting eq \ref{es24} from \ref{es1}, we get
\begin{multline}
    E_{ANM}-P_{ANM}=\overline{\epsilon}_F\left(\frac{1}{f}-\frac{5}{3f}+1 \right)+T_3\frac{1-\sigma}{\sigma+1}\left(\frac{\rho}{\rho_0}\right)^{\sigma}+\delta^2_x\left(S-\frac{L}{3}\right)\\
    =\overline{\epsilon}_F\left(1-\frac{2}{3f}\right)+T_3\frac{1-\sigma}{\sigma+1}\left(\frac{\rho}{\rho_0}\right)^{\sigma}+\delta^2_x\left(S-\frac{L}{3}\right)
    \label{es26}
\end{multline}
The third term of eq \ref{es26} can be determined from eq \ref{es25}, as 
\begin{equation}
    9T_3\frac{\sigma(\sigma-1)}{\sigma+1}\left(\frac{\rho}{\rho_0}\right)^{\sigma}= K_{ANM}-\overline{\epsilon}_F\left( \frac{10}{f} -12 \right)-\delta^2_x K_{sym}
    \label{es27}
\end{equation}
or equivalently 
\begin{equation}
    T_3\frac{\sigma(1-\sigma)}{\sigma+1}\left(\frac{\rho}{\rho_0}\right)^{\sigma}= -\frac{1}{9}\left[K_{ANM}-\overline{\epsilon}_F\left( \frac{10}{f} -12 \right)-\delta^2_x K_{sym}\right]
    \label{es28}
\end{equation}
By combining eq \ref{es26} with \ref{es28}, we get
\begin{multline}
    E_{ANM}-P_{ANM}=\overline{\epsilon}_F\left(1-\frac{2}{3f}\right)-\frac{1}{9}\left[K_{ANM}-\overline{\epsilon}_F\left( \frac{10}{f} -12 \right)-\delta^2_x K_{sym}\right]\\
    +\delta^2_x\left(S-\frac{L}{3}\right)=\overline{\epsilon}_F\left(1-\frac{2}{3f}-\frac{10}{9\sigma f}-\frac{12}{9\sigma}\right)-\frac{K_{ANM}}{9\sigma}\\
    +\delta^2_x \left(S-\frac{L}{3}+\frac{K_{sym}}{9\sigma}\right)
    \label{es29}
\end{multline}
Finally, the total EoS of NM at zero temperature, can be written as
\begin{equation}
      E_{NM}=P-\frac{K}{9\sigma}+\frac{\overline{\epsilon}_F}{9\sigma}\left(\frac{10-6\sigma}{f} +9\sigma -12\right)+\delta^2_x \left(S-\frac{L}{3}+\frac{K_{sym}}{9\sigma}\right)
    \label{es30}  
\end{equation}
The above formula holds for both SNM ($\delta_x = 0$) and ANM ($\delta_x \neq 0$), if the correct values of $P$ and $K$ are used.
\section{The EoS of Mean-Field Dynamics}
\label{s5}
\begin{figure}[!h]                                        
\centering
\includegraphics[height=12.0 cm]
{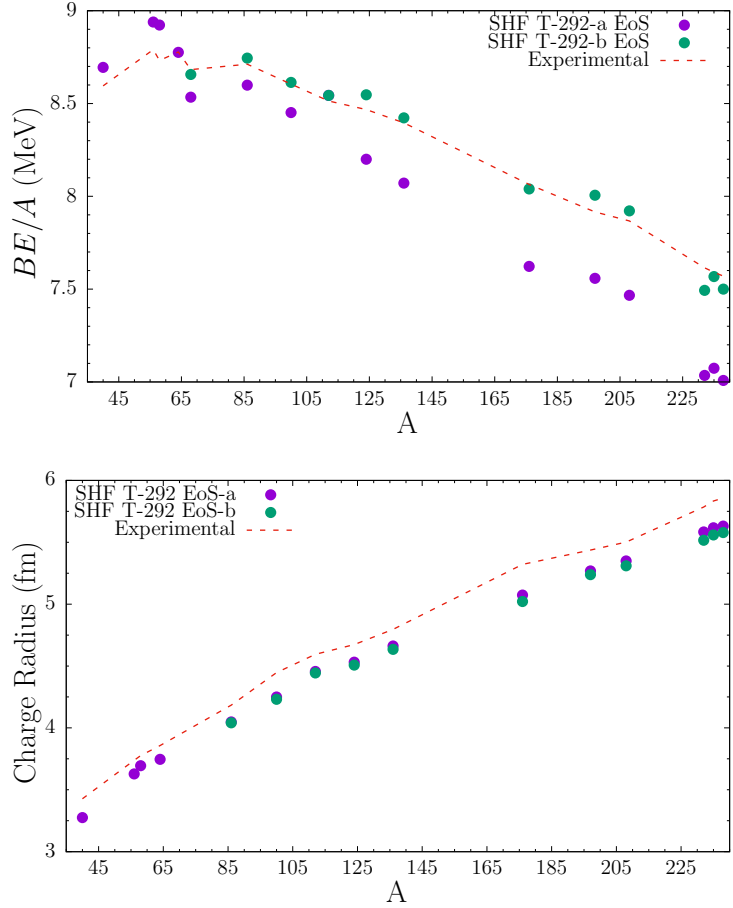} 
\caption{SHF calculations of binding energies per nucleon (top) and charge radii (bottom) of A=40-238 nucleim with different Coulomb contributions. The calculation were performed with the T-292 EoS with b=0.6 fm, $a_{sym}$=32 MeV and $m^*$=0.85m. The (a) calculation contains the full Coulombic contribution, while (b) accounts for the exchange term with a reduced Coulomb potential. The experimental data (dashed curve) were taken from \cite{Basdevant}(top) and \cite{Angeli2013}(bottom).}
\label{Fig3}
\end{figure}

The search of a universal EoS, i.e. an EoS that can describe properly all the nuclear degrees of freedom, requires us to postulate that the underline law of Nuclear Physics are independent of their mathematical model. Consequently, we assume that an EoS that can describe to a reasonable extent the properties of NM with within the Fermionic Dynamics framework (e.g. with the CoMD model) should describe the nuclear ground states with the Mean-Field approach (e.g. the SHF model), if is transformed  correctly. To that end, we tried to mathematically connect the EoS of CoMD and SHF. According to \cite{Vauth1971} the EoS of a Skyrme functional with parametrized density-dependent 3-body term and $x_1=x_2=x_3=0$ is:
\begin{equation}
    E_{SNM}=\overline{\epsilon}_F+\frac{3}{8}t_0\rho+\frac{1}{16}t_3\rho^{a+1}+\frac{3}{80}\left(3t_1+5t_2\right)\rho k_F^2
    \label{e43}
\end{equation}
where $\overline{\epsilon}_F$ is the mean Fermi energy with $m^*=m$, $k_F$ is the Fermi wavenumber and the lower case letter constants $t_1,t_2,t_3$ are the SHF potential parameters, different from the upper case letter parameters $T_1,T_3$ of the CoMD potential. Additionally, the effective mass is given by
\begin{equation}
    \frac{\hbar^2}{2m^*}=\frac{\hbar^2}{2m^*}+\frac{1}{16}\left(3t_1+5t_2\right)\rho
    \label{e44}
\end{equation}
By solving for $1/f$, we obtain eq \ref{e21}
\begin{equation}
    \frac{1}{f}=\frac{m}{m^*}=1+\frac{2m}{16\hbar}\left(3t_1+5t_2\right)\rho=1+\frac{m}{8\hbar}\left(3t_1+5t_2\right)\rho\equiv1+\frac{m}{8\hbar}\Theta_s\rho
    \label{e45}
\end{equation}
or equivalently 
\begin{equation}
\frac{1}{8}\left(3t_1+5t_2\right)\rho=\left(\frac{m}{m^*}-1\right)\frac{\hbar^2}{m}=\left(\frac{1}{m^*}-\frac{1}{m}\right)\hbar^2 
\label{e46}
\end{equation}
By inserting eq. \ref{e46} into eq. \ref{e43} we obtain
\begin{equation}
    E_{SNM}=\overline{\epsilon}_F+\frac{3}{8}t_0\rho+\frac{1}{16}t_3\rho^{a+1}+\frac{3}{10}\left(\frac{1}{m^*}-\frac{1}{m}\right)\hbar^2 k_F^2
    \label{e47}
\end{equation}
or equivalently
\begin{equation}
    E_{SNM}=\overline{\epsilon}_F-\frac{3}{5}\frac{\left(\hbar k_F\right)^2}{2m}+\frac{3}{8}t_0\rho+\frac{1}{16}t_3\rho^{a+1}+\frac{3}{5}\frac{\left(\hbar k_F\right)^2}{2m^*}
    \label{e48}
\end{equation}
By recognising the Fermi energy relation $\overline{\epsilon}_F=\frac{3}{5}\frac{\left(\hbar k_F\right)^2}{2m}$, the EoS of SNM with the SHF approach can be written as 
\begin{equation}
    E_{SNM}=\frac{\overline{\epsilon}_F}{f}+\frac{3}{8}t_0\rho+\frac{1}{16}t_3\rho^{a+1}
    \label{e49}
\end{equation}
By comparing eq \ref{e31} and \ref{e49} we obtain the relations between the constants $t_0, t_3, \alpha$ and $T_0, T_3, \sigma$
\begin{equation}
    \sigma=a+1
    \label{e50}
\end{equation}
\begin{equation}
    t_0=\frac{4}{3}\frac{T_0}{\rho_0}
    \label{e51}
\end{equation}
\begin{equation}
    t_3=\frac{16}{\sigma+1}\frac{T_3}{\rho_0^\sigma}
    \label{e52}
\end{equation}
To constrain the parameters $t_1, t_2$ and $x_0$ we utilise the symmetry energy (from ANM EoS) and surface thickness coefficients, as well as the effective mass parameter $\Theta_s$. For the effective mass we transform eq \ref{e21} as follows
\begin{equation}
    \Theta_s=\left(\frac{1}{f}-1\right)\frac{8\hbar^2}{m\rho}=3t_1+5t_2
    \label{e53}
\end{equation}
The symmetry energy coefficient is given by \cite{Vauth1971}
\begin{equation}
    a_{sym}=\frac{5}{9}\overline{\epsilon}_F-\frac{t_0}{4}\left(x_0+\frac{1}{2}\right)\rho-\frac{t_3}{16}\rho^{a+1}+\frac{1}{6}\rho k_F^2t_2
    \label{e54}
\end{equation}
The last term of eq \ref{e54} can be written as
\begin{equation}
    \frac{1}{6}\rho k_F^2t_2=\frac{3}{5}\frac{\left(\hbar k_F\right)^2}{2m}\frac{2m\rho t_2}{6\hbar^2}=\overline{\epsilon}_F\frac{m\rho t_2}{3\hbar^2}\frac{5}{3}
    \label{e55}
\end{equation}
By $t_2$ from eq \ref{e53} into eq \ref{e55}, we get
\begin{equation}
    \frac{1}{6}\rho k_F^2t_2=\overline{\epsilon}_F\frac{m\rho}{3\hbar^2}\frac{5}{3}\frac{\Theta_s-3t_1}{5}=\frac{\overline{\epsilon}_F\rho m}{9\hbar^2}\left[ \left(\frac{1}{f}-1\right)\frac{8\hbar^2}{m\rho}-3t_1\right]
    \label{e56}
\end{equation}
or equivalently
\begin{equation}
    \frac{1}{6}\rho k_F^2t_2=\frac{\overline{\epsilon}_F}{9}\left[ 8\left(\frac{1}{f}-1\right)-\frac{3t_1\rho m}{\hbar^2}\right]
    \label{e57}
\end{equation}
Combining eq \ref{e57} and \ref{e54}
\begin{equation}
    a_{sym}=\overline{\epsilon}_F\left[\frac{5}{9}+\frac{8}{9}\left(\frac{1}{f}-1\right)-\frac{t_1\rho m}{3\hbar^2}\right]-\frac{t_0}{4}\left(x_0+\frac{1}{2}\right)\rho-\frac{t_3}{16}\rho^{a+1}
    \label{e58}
\end{equation}
The first term of the above eq denoted as '$\Omega$' is:
\begin{equation}
    \Omega=\overline{\epsilon}_F\left[\frac{5}{9}+\frac{8}{9}\left(\frac{1}{f}-1\right)-\frac{t_1\rho m}{3\hbar^2}\right]=\frac{\overline{\epsilon}_F}{3}\left(\frac{8}{3f}-1-\frac{t_1\rho m}{\hbar^2}\right)
    \label{e59}
\end{equation}
The final equation we need to constrain the $t_1, t_2$ and $x_0$ parameters. According to \cite{Vauth1971}, nucleonic densities as a function radius can be fitted with the square of a hyperbolic tangent function. This fit has a surface thickness
\begin{equation}
    b=0.16\left[\frac{1}{\epsilon_F}\left(9t_1-5t_2\right)\rho\right]^{1/2} \rightarrow \left(\frac{b}{0.16}\right)^2=\frac{3}{5\overline{\epsilon}_F}\left(9t_1-5t_2\right)\rho
    \label{e60}
\end{equation}
By using eq \ref{e53} $t_2$ is given by
\begin{equation}
    t_2=\frac{1}{5}\left(\Theta_s-3t_1\right)
    \label{e61}
\end{equation}
By substituting eq \ref{e61} into eq \ref{e60}
\begin{equation}
   \left(\frac{b}{0.16}\right)^2=\frac{3}{5\overline{\epsilon}_F}\left(9t_1-\Theta_s+3t_1\right)\rho=\frac{3\rho}{5\overline{\epsilon}_F}\left(12t_1-\Theta_s\right)
   \label{e62}
\end{equation}
Then, solving with respect to $t_1$
\begin{equation}
    t_1=\frac{1}{12}\left[\frac{5}{3\rho}\overline{\epsilon}_F\left(\frac{b}{0.16}\right)^2+\Theta_s\right]
    \label{e63}
\end{equation}
Finally, $x_0$ is given by substituting eq \ref{e59} and \ref{e63} into \ref{e58}
\begin{equation}
    a_{sym}-\Omega+\frac{t_3}{16}\rho^\sigma=-\frac{t_0\rho}{4}\left(x_0+\frac{1}{2}\right)
    \label{e64}
\end{equation}
or equivalently 
\begin{equation}
    x_0=-\left(a_{sym}+\frac{t_3}{16}\rho^\sigma-\Omega\right)\frac{4}{t_0\rho}-\frac{1}{2}
    \label{e65}
\end{equation}
With eq \ref{e50}-\ref{e52}, \ref{e61}, \ref{e63} and \ref{e65} we properly transform the CoMD EoS of the previous sections into a compatible form within the framework of SHF. By using the code of \cite{CompNucPhys1}, we calculated the binding and shell energies, as well as the radii of several nuclides in the range A=40-238. Our results are plotted in fig \ref{Fig3}.
In fig \ref{Fig3}, we show the plot of BE/A (top) and charge rms radius (bottom) as a function A. The calculations were performed with the SHF-modified T-292 EoS. For the calculation of $t_1, t_2$ and $x_0$ parameters we used the values b=0.6 fm, $a_{sym}$=32 MeV and $m^*$=0.85m. We present two different calculations (a, purple and b green curve). The difference between the two is the contribution of the Coulomb potential. Both calculations take into account only the direct Coulomb term and calculation (a) accounts its full contribution. This makes nuclei underbound with respect to experimental data \cite{Basdevant}, especially heavier nuclei. To remedy this and approximately account for the exchange (stabilizing) contribution, we reduced the Coulomb contribution by 10\% for A=90-124 and by 20\% for A$>$124. This procedure results in calculation (b). Furthermore, both calculations appropriately reproduce the experimental \cite{Angeli2013} charge radii.
\section{Conclusion}
\label{s6}
The nuclear EoS lies in the heart of near-ground state dynamical properties of nuclei. In this work, we present an analytical derivation of the symmetric nuclear matter equations with effective mass equal to the free mass, in the framework of the CoMD model. Furthermore, we developed an approach for the EoS of the CoMD, that is consistent with a density dependent effective mass. Then the SNM equations are derived and solved for various semi-hard EoS with effective mass value $m^*=0.85m$. We plan to continue our study of the various low-energy dynamical phenomena with this and similar EoS and try to develop parameter sets for a 'soft' symmetry energy functional.\\
In addition to our study of the EoS for SNM, we extended our theory to include the symmetry energy and described the EoS of ANM. The equations for the asymmetric part of the EoS are derived similarly to the corresponding of the symmetric part. Using these equations we studied the symmetry energy of the EoS that are used in the CoMD model and constrained the value of the asymmetry parameter between $\pm 0.72$, for the NM. We also derived the total EoS of NM as a function of its thermodynamic parameters, the three body density parameter and the effective mass. \\
Finally we studied the ground state properties of several nuclei in the range A=40-238 with the SHF approach and the new T-292 EoS. A process of transforming a CoMD consistent to a SHF consistent EoS and vice versa was developed, with a mean field approximation. The SHF modified T-292 EoS with a reduced Coulomb potential reproduces properly the binding energies and charge radii of the aforementioned nuclei.


\begin{thebibliography}{99}


\bibitem{Takigawa} N. Takigawa, K. Washiyama, 'Fundamentals of Nuclear Physics', Springer Japan, (2017).

\bibitem{Ring} P. Ring, p. Schuck, 'The Nuclear N-Body Problem', Springer Berlin, Heidelberg, (1980).

\bibitem{Sk1958} T.H.R.Skyrme, Nuc. Phys. {\bf 9}, 4, 615-634 (1958).

\bibitem{Vauth1971} D. Vautherin, D.M. Brink, Phys Rev C {\bf 5}, 3 (1971).

\bibitem{Bon2001}  M. Papa, T. Maruyama, A. Bonasera
Phys. Rev. C {\bf 64}, 024612 (2001).

\bibitem{Papa2013} M Papa, J. Phys.: Conf. Ser. {\bf 420}, 012082 (2013).

\bibitem{Teo2021} T. Depastas, G.A. Souliotis, K. Palli, A. Bonasera, H. Zheng EPJ Web of Conferences {\bf 252}, 07003 (2021).


\bibitem{Cochet2004} B.Cochet, K. Bennaceur, P.Bonche, T.Duguet, J.Meyera, Nucl. Phys. A {\bf 731}, 34-40 (2004).


\bibitem{Zheng2014} G. Giuliani, H. Zheng, A. Bonasera, Prog. Part. Nuc. Phys. {\bf 76}, 116-164 (2014).



\bibitem{CompNucPhys1} K. Langanke-Joachim, A. MaruhnS. E. Koonin, Computational Nuclear Physics 1, Springer, Berlin, Heidelberg (1991).

\bibitem{Basdevant} J. L. Basdevant, J. Rich, M. Spiro, 'Fundamentals in nuclear physics: From nuclear structure to cosmology', New York: Springer, (2005).

\bibitem{Angeli2013} I. Angeli, K.P. Marinova, At. Data Nucl. Data Tables {\bf 99}, 69–95 (2013).


\end{thebibliography}
\end{document}